\documentclass{moriond}
\usepackage{printlen}
\usepackage{amssymb,amsmath}

\def\Journal#1#2#3#4{{#1} {\bf #2}, #3 (#4)}

\def\PRD{{\em Phys. Rev.} D}

\def\EPJC{{\em Eur. Phys. J.} C}
\def\JHEP{{\em JHEP}}

\def\be{\begin{equation}}
\def\ee{\end{equation}}
\def\bea{\begin{eqnarray}}
\def\eea{\end{eqnarray}}

\newcommand{\Photo}{}

\begin{document}
\vspace*{4cm}
\title{Constraining the real singlet extension of the SM: \\implications for vacuum stability}

\author{\underline{Moritz Bosse}$^{a}$, Gudrun Hiller$^{a,b}$, Daniel Litim$^{c}$, Fabio Maltoni$^{b,d,e}$, Michael J. Ramsey-Musolf$^{f,g,h}$, Simone Tentori$^{d}$, Guotao Xia$^{f,g}$}

\address{$^{a}$ TU Dortmund University, Department of Physics, Otto-Hahn-Str.~4, D-44221 Dortmund, Germany\\$^{b}$ Theoretical Physics Department, CERN, 1211 Geneva 23, Switzerland\\$^{c}$ Department of Physics and Astronomy, University of Sussex, Brighton, BN1 9QH, U.K.\\$^{d}$ Centre for Cosmology, Particle Physics and Phenomenology (CP3), Universit\'e Catholique de Louvain, B-1348 Louvain-la-Neuve, Belgium\\$^{e}$ Dipartimento di Fisica e Astronomia, Universit\`a di Bologna and INFN, Sezione di Bologna, Via Irnerio 46, 40126 Bologna, Italy\\$^{f}$ Tsung-Dao Lee Institute \& School of Physics and Astronomy, Shanghai Jiao Tong University, Shanghai 200240, China\\$^{g}$ Shanghai Key Laboratory for Particle Physics and Cosmology, Key Laboratory for Particle Astrophysics and Cosmology (MOE), Shanghai Jiao Tong University, Shanghai 200240, China\\$^{h}$ Kellogg Radiation Laboratory, California Institute of Technology, Pasadena, CA 91125, USA}

\maketitle
\abstracts{
Amongst the simplest extensions of the Standard Model is the addition of a real singlet scalar field with profound phenomenological consequences. The singlet can catalyze a strong first-order electroweak phase transition, necessary for successful electroweak baryogenesis. Furthermore, the current prediction of a metastable Higgs vacuum can be lifted towards absolute stability. 
While current data mainly constrain the scalar mixing angle, future measurements of the Higgs self-coupling and direct searches for additional scalar states will probe much larger parts of the viable parameter space.
We report on a combined study of theoretical and experimental constraints on the real singlet extension, highlighting here in particular its implications for vacuum stability.
}

\section{Introduction}

The addition of a real scalar singlet to the Standard Model (SM) is among its simplest and most motivated extensions. Coupled to the Higgs through the portal interaction, the singlet can have important consequences for collider phenomenology, the electroweak phase transition, and the stability of the Higgs potential at high energies. In particular, it can strengthen the electroweak phase transition and shift the metastable SM vacuum towards absolute stability.

In this contribution, we focus on the implications for vacuum stability. Owing to the near-criticality of the SM, even small singlet-induced effects in the renormalization group evolution and in threshold corrections can qualitatively alter the fate of the electroweak vacuum. 
A comprehensive analysis of theoretical and experimental constraints on the model is presented in Ref.~\cite{Bosse:2026}, on which this contribution is based. After reviewing the current status of vacuum stability in the SM, we discuss the impact of the singlet through the so-called singlet portal mechanism. 
We conclude with a benchmark illustrating the interplay of theoretical constraints, Run-2 searches, and high-luminosity LHC (HL-LHC) projections.

\section{The fate of the electroweak vacuum}

The SM is known to be near-critical, with the Higgs quartic coupling $\lambda$ running to negative values at high energies, indicating a metastable vacuum. After a decade of precision measurements, the remaining uncertainty on this conclusion is dominated by the top quark pole mass, which is currently measured to be $m_t = 172.4 \pm 0.7$ GeV~\cite{ParticleDataGroup:2024cfk} and the strong coupling constant $\alpha_s$, which is measured to be $\alpha_s(m_Z) = 0.1180 \pm 0.0009$~\cite{ParticleDataGroup:2024cfk}.

Theory calculations have reached a very high level of precision, with the renormalization group equations (RGEs) and the effective potential known to full three-loop order and the low-scale matching to $\overline{\mathrm{MS}}$ couplings known to two-loop order. 

\begin{figure}
       \centering
       \includegraphics[width=0.6\textwidth]{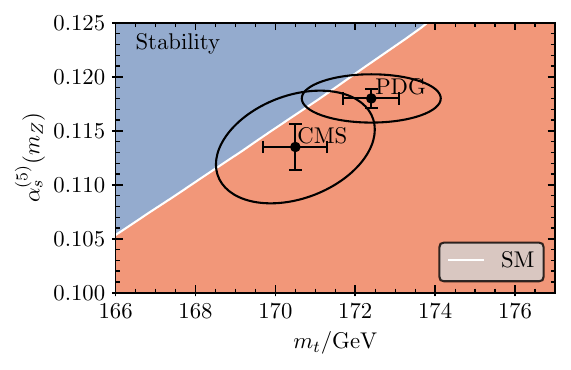}
       \caption{Vacuum stability in the SM as a function of the top quark mass $m_t$ and the strong coupling constant $\alpha_s$. The blue region indicates absolute stability while red is the combination of metastability/instability. Shown are the central values and 2\,$\sigma$ uncertainty bands corresponding to the current PDG values \protect\cite{ParticleDataGroup:2024cfk}
       and to a previous correlated CMS analysis\protect\cite{CMS:2019esx}. Plot adapted from Ref.~\protect\cite{Bosse:2026}.}
       \label{fig:vacuum_stability}
\end{figure}

The resulting stability phase diagram is shown in Fig.~\ref{fig:vacuum_stability}. The present measurements lie close to the boundary between absolute stability and metastability/instability, illustrating the near-criticality of the SM. Although stability cannot be ruled out at the current level of precision, this near-criticality makes vacuum stability an interesting guide for well-motivated extensions of the SM. A detailed discussion on the stability in the SM can be found in Ref.~\cite{Hiller:2024zjp}.

\section{The real singlet extension of the SM}

The real singlet extension of the Standard Model, also referred to as the Real Singlet Model (xSM), extends the SM by a real scalar field $S$ which is a singlet under the SM gauge groups. The most general renormalizable scalar potential can be written as 
\be
V(H,S) = -\mu_H^2 |H|^2 + \lambda |H|^4 + \frac{a_1}{2} |H|^2 S + \frac{a_2}{2} |H|^2 S^2 + b_1 S + \frac{b_2}{2} S^2 + \frac{b_3}{3} S^3 + \frac{b_4}{4} S^4.
\ee
For a detailed discussion of the model and conventions, see e.g. Ref.~\cite{Bosse:2026}. 

The phenomenology is mainly controlled by the Higgs--singlet mixing angle $\theta$, the mass of the additional scalar state, and the portal coupling $a_2$. Mixing induces a universal rescaling of the observed Higgs couplings, while the second scalar can be probed in resonant searches. In addition, the model allows for sizable modifications of the trilinear Higgs self-coupling and can realize a strong first-order electroweak phase transition.

\subsection*{Vacuum stability through the singlet portal}

The presence of the singlet modifies the RGEs of the Higgs quartic coupling $\lambda$ and can give rise to threshold corrections at the scale of the singlet mass. Both effects can potentially (de)stabilize the vacuum up to the Planck scale. 
For the former, only the marginal couplings enter the relevant RGEs at large field values $h \gg v$, where the metastability in the SM arises. The relevant new contributions are therefore the portal $a_2$ and the singlet quartic $b_4$. At leading order 
\be
\beta_\lambda = \beta_\lambda^{\mathrm{SM}} + \frac{1}{(4\pi)^2} \frac{1}{2} a_2^2,
\ee
the portal coupling $a_2$ gives a positive contribution to the running of $\lambda$, irrespective of its sign, and can therefore stabilize the vacuum. In contrast, dependence on the singlet quartic $b_4$ enters only at three-loop order and is therefore highly subleading. Nevertheless, $b_4$ significantly affects the running of $a_2$ and hence indirectly influences vacuum stability. While this indirect effect can be crucial at larger $b_4$ values, we find it to be negligible in the range $0 \leq b_4 \lesssim 0.3$ \cite{Bosse:2026}.

New instabilities may arise from the presence of the new singlet direction in the potential. Necessarily, the potential must be bounded from below in all field directions, requiring the following conditions on the quartic couplings:
\be
\lambda > 0, \quad b_4 > 0, \quad a_2 + 2 \sqrt{\lambda b_4}> 0.
\ee

\begin{figure}
       \centering
       \includegraphics[width=\textwidth]{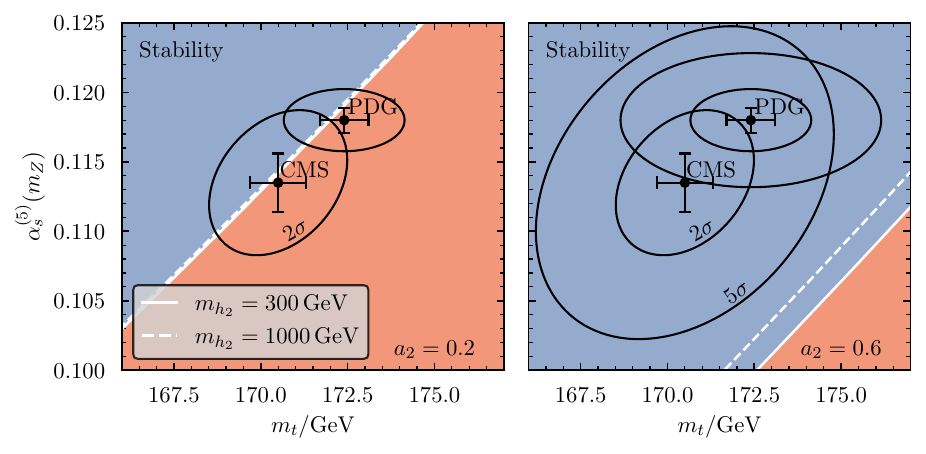}
       \caption{Stability phase diagram in the xSM for different values of the portal coupling $a_2 = 0.2, 0.6$ and masses $m_{h_2} = 300, 1000\,\mathrm{GeV}$ of the additional scalar state. For simplicity, we fix the singlet quartic $b_4 \simeq 0$ at the electroweak scale. Blue denotes absolute stability, while red is the combination of metastable and unstable regions. Measurements shown as in Fig.~\ref{fig:vacuum_stability} with an additional 5\,$\sigma$ uncertainty band for the $a_2 = 0.6$ (right) benchmark. Plot adapted from Ref.~\protect\cite{Bosse:2026}.
       }
       \label{fig:vacuum_stability_rxsm}
\end{figure}

The mixed field direction requires negative values of the portal coupling $a_2$ to be sufficiently small in magnitude. This effect is further enhanced through the RG. In fact, it was shown that without additional threshold/mixing corrections negative values of $a_2$ are not compatible with absolute stability up to the Planck scale~\cite{Hiller:2024zjp}. 
This can be understood as another peculiar consequence of the SM being close to criticality.
By including the threshold/mixing corrections, the stability of the vacuum can be lifted towards absolute stability for a wide range of parameters, including negative values of $a_2$~\cite{Bosse:2026,Elias-Miro:2012eoi}. 
Figure~\ref{fig:vacuum_stability_rxsm} illustrates this stabilizing effect in the limit of vanishing Higgs--singlet mixing, with threshold effects reduced by matching at the new scalar mass scale $m_{h_2}$ and $b_4\simeq0$ at the electroweak scale \footnote{This choice does not induce a runaway direction, as a positive $b_4$ is generated radiatively.}.
Above that scale, the RG evolution includes the singlet contributions and is performed using three-loop gauge and two-loop $\beta$ functions otherwise.
Already moderate portal couplings substantially enlarge the stable region: $a_2=0.2$ moves it close to the current central values, while $a_2=0.6$ gives stability at the 5\,$\sigma$ level.

\subsection*{Combination of theoretical and experimental constraints}

\begin{figure}
       \centering
       \includegraphics[width=0.45\textwidth]{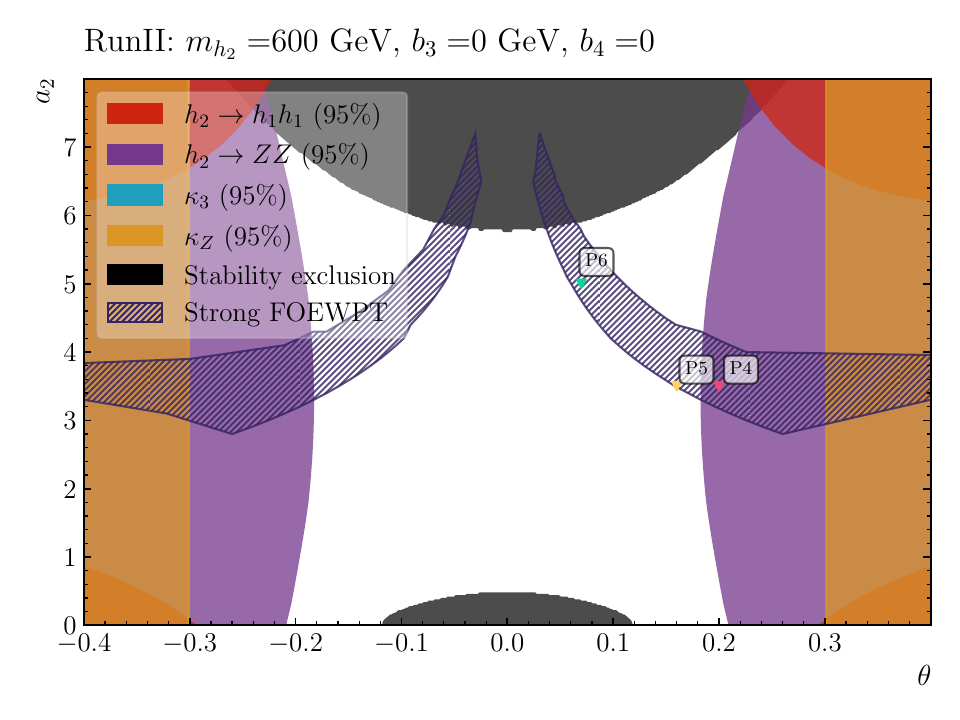}
       \includegraphics[width=0.45\textwidth]{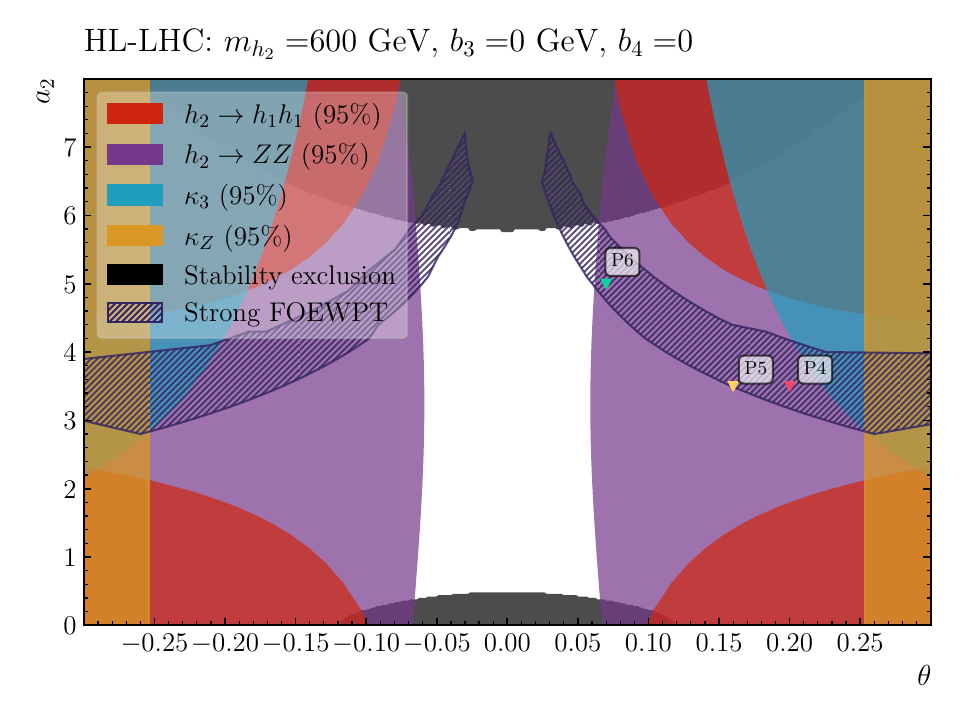}

       \caption{Combined Run-2 constraints (left) and HL-LHC projections (right) for $m_{h_2}=600\,\mathrm{GeV}$, with electroweak-scale input values $b_3=0\,\mathrm{GeV}$ and $b_4=0$, in the $\theta-a_2$ plane. Exclusion regions are shown at 95\% C.L.; resonant searches are evaluated in the Narrow Width Approximation. The dashed region indicates a strong first-order electroweak phase transition. The black region is excluded by vacuum stability up to the Planck scale and the absence of sub-Planckian Landau poles. Plots adapted from Ref.~\protect\cite{Bosse:2026}.}
       \label{fig:combined_constraints}
\end{figure}

We finally illustrate how the vacuum-stability requirement complements present and future collider probes. Figure~\ref{fig:combined_constraints} shows the combined constraints in the $\theta-a_2$ plane for the benchmark $m_{h_2}=600\,\mathrm{GeV}$, with electroweak-scale input values $b_3=0\,\mathrm{GeV}$ and $b_4=0$. Current Run-2 sensitivity is mainly driven by resonant searches, in particular $h_2\to h_1h_1$ and $h_2\to ZZ$, together with Higgs signal-strength measurements. These constraints primarily restrict the mixing angle $\theta$. By contrast, the requirement of absolute vacuum stability up to the Planck scale, together with the absence of sub-Planckian Landau poles, mostly constrains the portal direction $a_2$.

The benchmark therefore displays a clear complementarity between experiment and theory. Regions compatible with a strong first-order electroweak phase transition extend to sizable portal couplings, which also tend to lift the metastable SM vacuum towards absolute stability. The HL-LHC is projected to probe most of this region for the benchmark shown. Measurements of the Higgs self-coupling provide additional sensitivity, especially where small mixing suppresses the reach of direct searches and Higgs signal-strength measurements.

\section{Summary}

The real singlet extension provides a minimal setting in which collider phenomenology, the electroweak phase transition, and high-scale vacuum stability are closely connected. The singlet portal coupling gives a positive contribution to the running of the Higgs quartic coupling, while threshold and mixing effects can further lift the metastable SM vacuum towards absolute stability. Moderate portal couplings can therefore substantially enlarge the stable region in the $(m_t,\alpha_s)$ plane. Future measurements at the HL-LHC, including the Higgs self-coupling, will probe much of the viable parameter space where a strong first-order electroweak phase transition and improved vacuum stability can be realized simultaneously.

\section*{Acknowledgments}

M.B. thanks his collaborators and Tom Steudtner for helpful discussions, and the organizers of the 60$^{\mathrm{th}}$ Rencontres de Moriond for the invitation to present this work.

\end{document}